# More on the Evidence for a Bubble Universe with a Mass ~$10^{21}$ M$_\odot$

Michael J. Longo[1]

Department of Physics, University of Michigan, Ann Arbor, MI 48109, USA

This is a very informal report that gives further details on the evidence for a bubble universe based on an anomaly in the angular distribution of quasar magnitudes that was presented in a short paper in arXiv:1202.4433. This report addresses some concerns of two reviewers. It is meant to be read in conjunction with 1202.4433. There is very little overlap between the two articles. This extended discussion is, by necessity, somewhat more technical in nature.

I am grateful for the reviewers' comments that forced me to understand these issues more thoroughly.

### REFEREE I

The main concern of the first reviewer is that the quasar sample is not uniform in that there are regions of the SDSS survey area where the selection algorithm was a bit different or the "serendipity" branch of the algorithm got more targets, etc. This reviewer also suggested that only those quasars identified by the "uniform selection" flag, which indicates that the object was a primary quasar target with the final selection algorithm given by Richards et al. [4], should be used. This selection includes less than half of the total quasar candidates in the catalog.

My reply below is quite lengthy so let me first summarize it. This reviewer makes a good point. However, it begs the question as to why the selection algorithm, <u>which contains no explicit dependence on right ascension and declination</u>, will cause the peculiar systematic dependence of the quasar magnitudes on RA, $\delta$, and redshift that is observed. Therefore the only way a systematic dependence on RA, $\delta$ could come about is through a large systematic bias that extends over a contiguous region of the sky ~±30° wide. The SDSS campaigns for the skirt and cap region surveys were done in drift scans with consistent exposures over each patch, as described in (i) below.

In the following, I confine my remarks to quasars with $z < 2.2$ that were used in the paper.

The "uniform" flag specifically finds only quasars with corrected $I$ magnitudes less than 19.1. This removes all of the fainter quasars that can have magnitudes up to about 21.5, and, as I discuss in detail below, this is where the evidence for the gravitational lens comes from.

The uniform flag was not meant to select a sample that was "uniform" in angle. According to Richards et al. [4], the purpose of the target selection algorithm was to study "the evolution of the quasar luminosity function and …. the spatial clustering of quasars as a function of <u>redshift.</u>

---

[1] email: mlongo@umich.edu

These studies require the assembly of a large sample of quasars covering a broad range of redshift and chosen with well defined, <u>uniform</u> selection criteria" (emphases added). The quasars were selected via their nonstellar colors in *UGRIZ* photometry; the criteria do not contain their position on the sky. Except for possibly a very few manually selected serendipity candidates, this category was also based on their lying outside the stellar locus in color space. Overall, there were less than 1000 candidates for which either the FIRST or ROSAT categories were the sole selection. The FIRST coverage is very similar to that of SDSS, and ROSAT is an all-sky catalog. Therefore these other categories can not be the source of the angular dependence.

In my analysis I had used objects with spectroscopic classification (specclass) = QSO or HIZ_QSO. I have repeated the analysis with a more transparent selection that was based on the Target Selection flags. This analysis chose quasar candidates in the skirt or cap regions and excluded luminous red galaxies (LRG's). The "uniform" flag was not specifically required, so that fainter quasars were included. It gave almost identical results to the analysis in the paper.

As seen in Fig. 4 of the paper, the brightness enhancement, which I will refer to simply as the "hotspot", occurs roughly for $z \gtrsim 0.5$, $160° < RA < 240°$ and from $\delta = 40°$ down to the lowest declinations covered by SDSS, about $-5°$. I attribute this brightness enhancement to the effects of a massive gravitational lens centered at $(RA, \delta) \sim (195°, 0°)$ and subtending an angular range $\sim \pm 30°$. In most of the following plots I look at $\delta < 40°$ to show the effect of the lens on the RA distributions.

The SDSS galaxy survey data were taken in the same exposures as the quasar data. As I discuss in detail below, the galaxy and LRG magnitude distributions, as well as the quasars with $z < 0.5$, do not show the systematic dependence of intensities on RA, $\delta$ that is shown by the quasars for $z \gtrsim 0.5$, so that the quasar brightness enhancement cannot be due to a systematic variation in the SDSS survey depth. On the other hand, all of the available evidence from quasars, LRG's, galaxy distributions, and CMB is consistent with the gravitational lens interpretation, as I discuss here and in the paper.

I first discuss how a large gravitational lens would show up in the quasar distributions in greater detail than I could in the letter article.

In Fig. 1a, I show a scatter plot of green magnitudes vs. RA for $\delta < 40°$ and $1.0 < z < 2.2$. The peculiar steplike structure in the magnitude distribution is due to branches in the selection algorithm that make angle-independent cuts at corrected infrared magnitudes of about 19 and 20 mag, depending on redshift range. There does seem to be a magnitude enhancement (shift to lower magnitudes) between about 160° and 240° that is consistent with the gravitational lens scenario. Its angular extent is more clearly visualized in Fig. 4 of the paper.

The effect of the "uniform" selection is shown in Fig. 1b. By design, it only passes quasars with green magnitudes below about 19.7. However, it then removes most of the evidence for the gravitational lens enhancement.

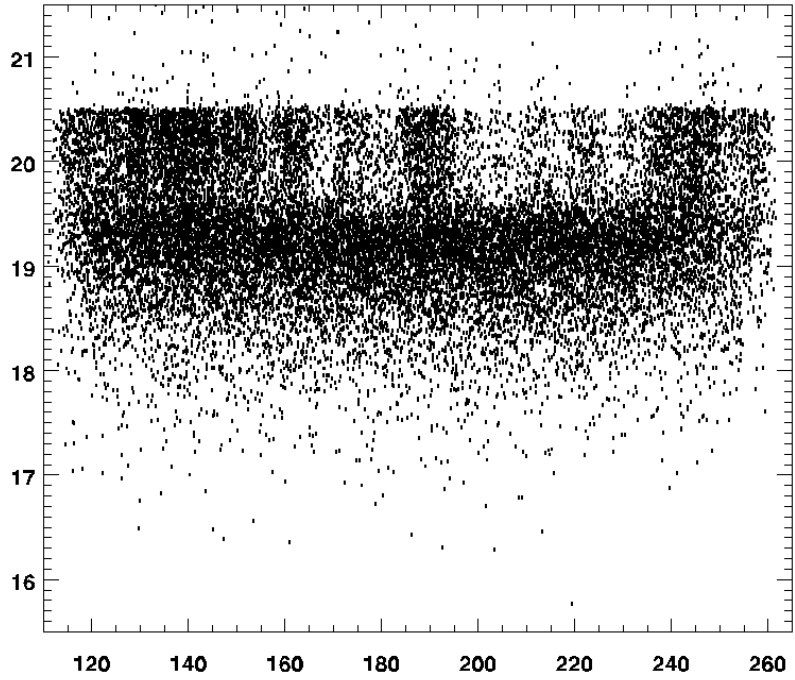
Figure 1a – Green magnitude vs. RA (degrees) for $\delta < 40°$ and $1.0 < z < 2.2$.

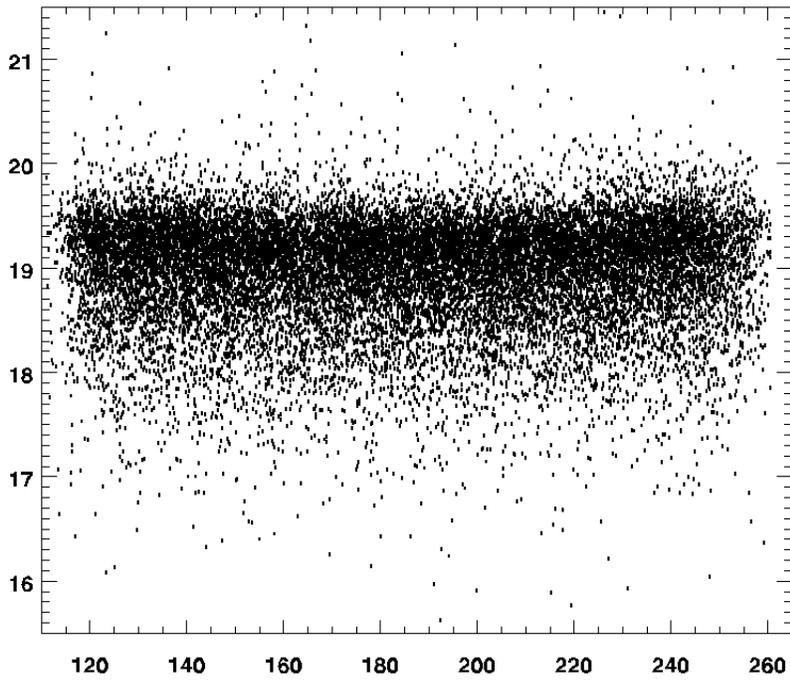
Figure 1b – Green magnitude vs. RA (degrees) for $\delta < 40°$ and $1.0 < z < 2.2$ with "Uniform" selection. The Uniform selection includes only candidates with extinction-corrected infrared magnitudes $\leq 19.1$.

One might expect this enhancement to also appear at fainter magnitudes in the 160°–240° range, but because of the steep falloff of quasars with increasing magnitude, this is not apparent in Fig. 1a. One would also expect a corresponding increase in brighter quasars behind the lens, but this is again not obvious in Fig. 1 because of the high density of points there. It can be seen in Fig. 2 that compares the magnitude distribution for the hotspot and elsewhere.

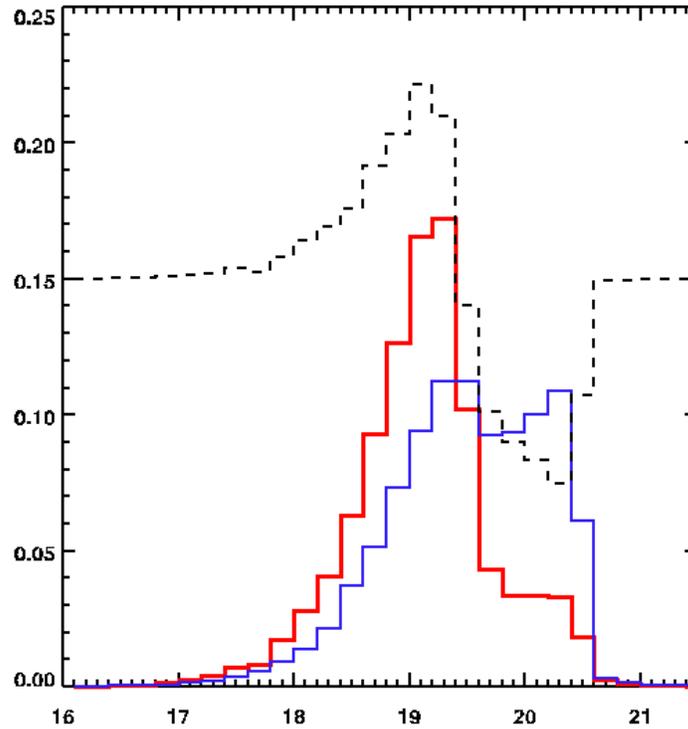

Figure 2 – Number vs. Green magnitude for hotspot (thicker red) and for remainder (blue), normalized to equal areas under the curve. The dashed curve is their difference shifted upward by 0.15. In the hotspot the magnitudes are significantly brighter.

The shift toward brighter magnitudes in the hotspot is apparent in Fig. 2. The same shift is observed for all the filter bands. The "shelf" around 20 mag is due to the steplike structure seen in Fig. 1. There is a sharp cutoff beyond 20.6 mag because the selection algorithms fail to find quasars there.

This behavior described above is exactly that expected if the hotspot is caused by gravitational lensing by a large overdense region. It is not easily explained by an angle dependent selection criterion that preferentially selects brighter objects over a cone with a half angle ~30° on the sky that is centered at approx. $(\alpha, \delta) = (195°, 0°)$.

Another characteristic that would be expected for a gravitational lens enhancement would be an apparent <u>decrease</u> in the number density of objects behind the lens due to the smaller solid

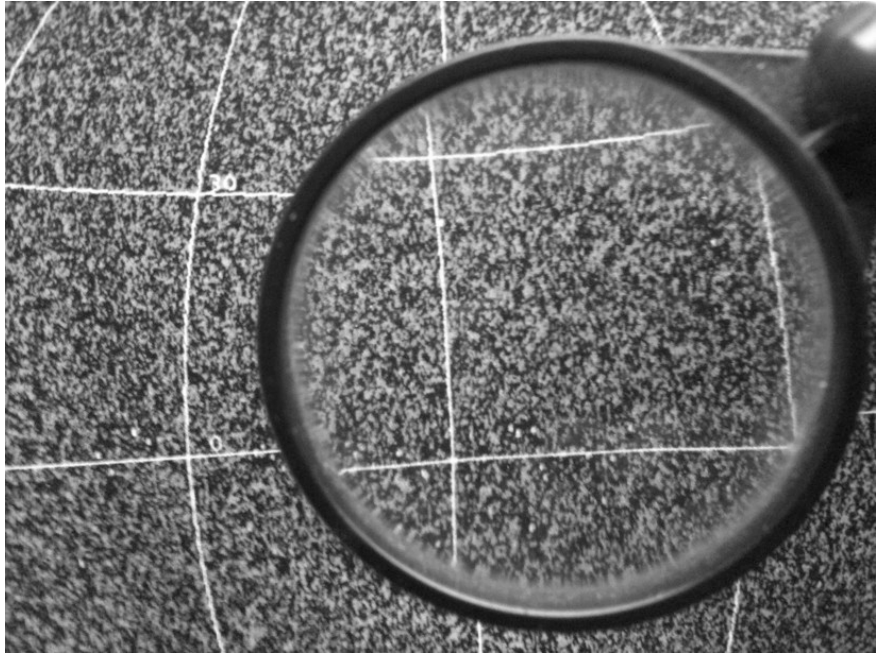
Figure 3–Analog simulation of the gravitational lens. Note the reduced solid angle seen behind the lens.

angle viewed behind the lens compared to outside. In the development of the gravitational lens model in the paper, this decrease in density is due to the same angular compression that causes the brightness enhancement. This happens for a magnifying glass or telescope and is demonstrated by an analog simulation in Fig. 3. The magnifying glass covers approximately the same angular region on the sky as the putative gravitational lens. The lens model therefore predicts a <u>lower</u> number density of quasars in the hotspot. Figure 4 shows a histogram of the number of quasars vs. RA for $\delta<30°$; there is a significantly lower density between 160° and 240° as anticipated. Note that if, as the reviewer suggests, the magnitude enhancement in the hotspot is due to a larger depth survey in that region, <u>more</u> quasars would be expected there, especially at fainter magnitudes.

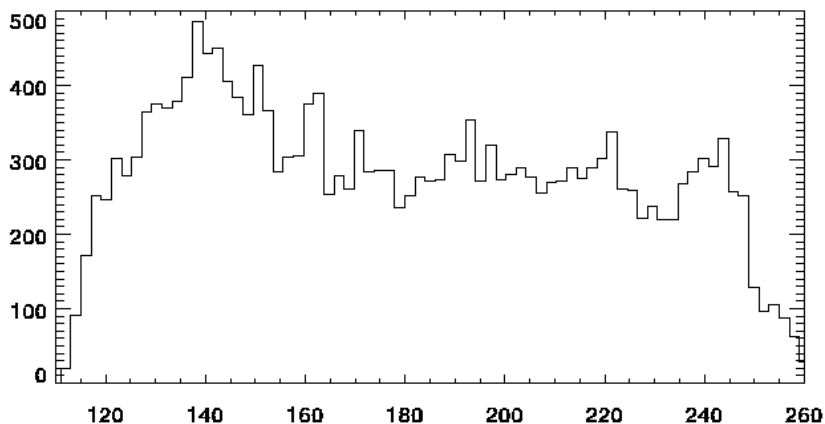
Figure 4 – Histogram of number of <u>quasars</u> vs. RA (degrees) for $-10° < \delta < 30°$, $0.6 < z < 2.2$

In addition, there are several other threads of evidence that the brightness enhancement is not due to a deeper or more sensitive survey in the angular region of the hotspot.

(i) The bulk of the SDSS spectroscopy was done with a drift scan with contiguous exposures along stripes in right ascension and declination that are about 2.5° wide, as described in York et al. (2000). The exposures were taken until a prescribed signal/noise ratio was achieved at corrected *G* magnitudes of 20.2 and *I* magnitudes of 19.9 was achieved. Thus the exposure-to-exposure variation in depth would be small and there would be no correlation of exposure depths between adjacent stripes or over large regions in $(\alpha, \delta)$ such as those observed that appear over a contiguous region ~50° wide on the sky.

(ii) The quasar spectroscopy data were taken in the same exposures as all the other star, galaxy, LRG, … samples. Any variation in the depth of the quasar survey would show up in the distributions for the others. A deeper survey would be most apparent in the fainter magnitudes. In Fig. 5, I show the number of galaxies vs. RA for Gmag > 18.5, similar to Fig. 4 for the quasars. Despite the very good statistics, there is no sign of the deficit above 160° observed for the quasars. In the $(\alpha, \delta)$ scatter plot corresponding to Fig. 4 of the paper, there is no sign of the hotspot in the galaxy angular distribution, while the well-known filamentary structures are visible.

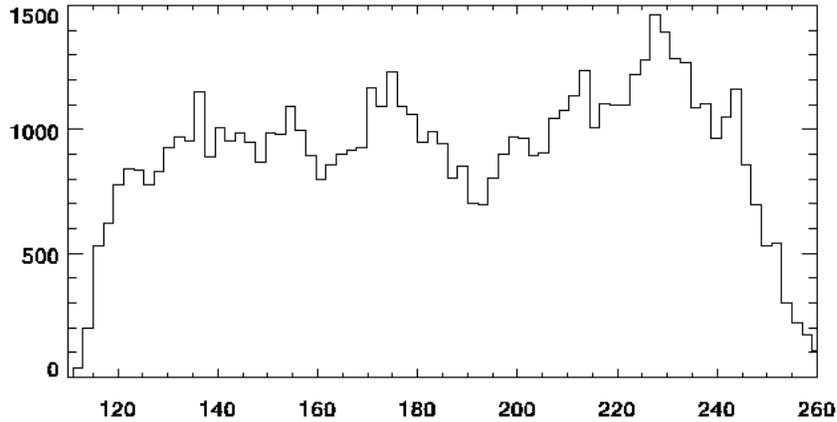

Figure 5– Histogram of number of galaxies vs. RA (degrees) for –10°<$\delta$<30°, 0.02 < *z* <0.2, Gmag > 18.5

(iii) If the depth of the survey varies systematically with $(\alpha, \delta, z)$, the brightest objects should show the same behavior as the quasars. In Fig. 6, I show the RA distribution for all SDSS objects with –10°<$\delta$<30°, 0.2 < *z* <0.35, Gmag > 18.5. This selection is dominated by LRG's. Again, there is no sign of the deficit between 160° and 240° that is shown by the quasars.

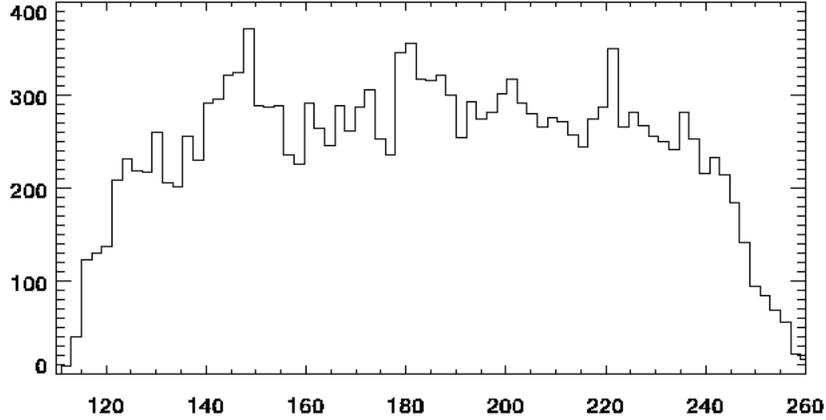
Figure 6– Histogram of <u>bright</u> <u>objects</u> (mostly LRGs) vs. RA for −10°<δ<30°, 0.2 < z <0.35, Gmag > 18.5

(iv) As seen in Figs. 2 and 3 of the paper, the brightness enhancement of the lens appears to be confined to $z \gtrsim 0.5$. If the reason for the apparent enhancement were due to a deeper survey in the hotspot region, the fainter quasars with $z < 0.5$ should show a surplus for RAs between 160° and 240°. In Fig. 7, I show the RA distribution for all SDSS quasars with −10° < δ <30°, 0.2 < z <0.5, and Gmag > 18.5. There is no sign of any surplus between 160° and 240° that would be expected if the quasar brightness enhancement were due to a deeper survey there. There is also no sign of the deficit that the quasars show for $z > 0.5$ behind the lens.

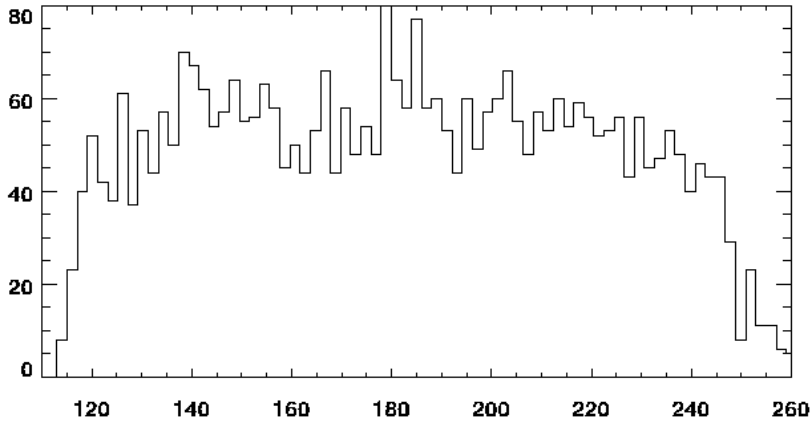
Figure 7– Histogram of number of quasars with 0.2 < z <0.5 and Gmag > 18.5 vs. RA for −10°<δ<30°.

(v) The magnitude shifts that I attribute to the lens are actually quite large. From the contours in Fig. 4 of the paper, they appear to be $\gtrsim 0.2$ mag between the hotspot and its surroundings. I believe this corresponds to an observation time that is systematically longer by approx. 20% for the entire hotspot region, which is not consistent with the survey strategy discussed in (i) above.

(vi) The hotspot is centered at Galactic latitude ~63°, and so it is far away from the plane of the Galaxy. Thus, the density of stars in that region is low and uniform in angle. Corrections in the magnitudes for galactic extinction were made and are very small.

(vii) As discussed in Richards et al. [4], the color magnitude locus used by the code is fixed by definition, any shift as a function of position on the sky would have the effect of greatly worsening the target selection efficiency in certain areas of sky. They tested this possibility by examining its parameters as a function of position on the sky, and found that the position of the color magnitude locus was consistent within the errors of their photometric calibration. No systematic trend with position on the sky was seen.

(vii) The surveys in the Southern Galactic cap region (right ascensions between 0° and 60° and between 300° and 360°) were done piecemeal in stripes and patches (SEGUE). These show no sign of angle dependent magnitude fluctuations. (See, for example, Figs. 2 and 3 in the paper.)

I believe the above discussion plus the abbreviated discussion in the paper give ample evidence that the quasar brightness enhancement is real and can be explained by the gravitational lensing of a massive bubble with $M_{lens}$ ~ $10^{21}$ M$_\odot$, a lens radius ~350 Mpc, and with the lens subtending an angle ~ ±30° on the sky. Of course, it is not possible to put all of this information in a short paper.

REFEREE II

The main concern of the second reviewer seems to be the possibility that the shifts in colors are actually due to a large-scale structure such as a filament or wall. However, the discovery of such a coherent large-scale structure that extends at least from $z=0.5$ to $z=2.2$ and subtends over 60° on the sky would be an even more remarkable discovery than a bubble universe. The quasar brightness enhancement extends over a distance scale $\sim 5 \times 10^9$ pc, while the largest known structures, superclusters, are ~100 times smaller. Such a large structure cannot be causally connected and would falsify the Cosmological Principle, as well as present severe difficulties for general relativity and would challenge the inflation paradigm. Attributing the observed anomalies in the quasar intensities to a massive gravitational lens seems to be the least radical explanation of these puzzling phenomena.

I know of no other physical models that could explain such a large scale, systematic intensity enhancement. Certainly any such model would contradict the Cosmological Principle in an irreparable way. At least the large gravitational lens explanation posits (perhaps) only one exception.

Quasars, by no means, can be considered standard candles. There is a large variation in magnitudes, even for quasars at approximately the same redshift. However, the Cosmological Principle requires their distributions to be homogeneous when averaged over sufficiently large volumes of space. Quasars offer our farthest reaching probe of the Universe except for the CMB. If the principle doesn't work for them, it is moot.

The second reviewer also questions the uniformity of the quasar sample. I discuss this question at great length in my reply to Referee I above. There is no basis for invoking nonuniformity as an explanation of these systematic effects that extend over such a large region in space.

Many of the difficulties that this reviewer sees in the manuscript are a result of the need to cram a lot of information into a short letter article. It was necessary to show the lens model predictions prematurely as dashed curves in Fig. 3 (left) simply to save space. The brightness enhancement appears as a downward shift in the dashed curves. Their vertical position was set by eye so that they are centered vertically around $\langle\Delta U\rangle \approx 0$. The apparent discontinuities at 140° and 240° reflect the angular extent of the presumed lens. In this simple multiple thin lens, small-angle, treatment, the edges are sharply defined, as in Fig. 3 above. I know of no treatment of a <u>large</u> gravitational lens in the literature, so a simplified model is justified. I will have to leave to the experts a more complete treatment without these approximations and with the presumed expansion of the bubble since its birth, its density profile, as well as gravitational time delays and other general relativistic effects taken into account. (See, for example, Johnson, Peiris, and Lehner, arXv:1112.4487v2, for a full general relativistic treatment of bubble collisions.) Because the gravitational lens covers such a large part of the survey region it's hard to appreciate what is happening outside of it. The portions of the survey at $\delta > 40°$ and RA ~ 0° do not seem to show the brightness enhancement or the depletion in number density described in my reply to Ref. I. To give some information on different filter bands, I show $U$ magnitude shifts in Figs 2 and 3 and *G* magnitude shifts in Fig. 4. I also show different $\delta$ slices in Figs. 2 and 3 to give more information on the $\delta$ variation.

The other (perhaps related) evidence for anomalies in roughly the same region, which I mention briefly in the Discussion section, is well described in Refs. [6] through [14]. There are already hundreds of articles on possible anomalies that tend to be in that portion of the sky. These generally have large uncertainties in their angular coordinates. The CMB dipole, though apparently well measured, has some uncertainty because of the uncertainty in how much of it is attributed to our proper motion. Antoniou and Perivolaropoulos [13] give a summary of the CMB, velocity flow, and supernovae evidence for anomalies and show error ellipses.

I appreciate most of the concerns of the second referee. I have made revisions in the paper to address some of them to the extent possible in a letter. I believe the paper already gives more detail than the average reader would care to read. There is a long tradition of first publishing important new results in brief form in a short article, followed by a more lengthy discussion in a regular article. This seems to be the appropriate course of action in this case. The follow up article will contain much of the material discussed in these replies and can contain more information on the filter bands to demonstrate that the effects are indeed achromatic.